\documentstyle[psfig]{elsart}

\def\gray{$\gamma$-ray}
\def\grays{$\gamma$-rays}

\def\apj{Astrophys. J.}
\def\apjl{Astrophys. J. (Lett.)}

\def\mic{$\mu$m}

\def\mqg{M_{\rm QG}}
\def\mpl{M_{\rm Planck}}

\def\epr{e-print}

\begin{document}

\begin{frontmatter}

\title{Tests of Quantum Gravity and Large Extra Dimensions Models using 
High Energy Gamma Ray Observations}

\thanks[talk]{Invited Talk Presented at the 2nd VERITAS Symposium on TeV 
Astrophysics 
of Extragalactic Sources, Chicago, IL, April 2003; to be published in
the Proceedings.}

\author{F.W. Stecker}

\address{Laboratory for High Energy Astrophysics, NASA Goddard Space Flight 
Center, Greenbelt, MD 20771, USA}

\begin{abstract}

Observations of the multi-TeV spectra of the nearby BL objects Mkn 421 and Mkn 
501 exhibit the high energy cutoffs predicted to be the result of 
intergalactic annihilation interactions, primarily with infrared photons 
having a flux level as determined by various astronomical observations. 
After correction for this absorption effect, the 
derived intrinsic spectra of these multi-TeV sources can be explained within 
the framework of simple synchrotron self-Compton emission
models. Stecker and Glashow have shown that the existence of such 
annihilations {\it via} electron-positron pair production interactions up to 
an energy of 20 TeV puts strong constraints on Lorentz invariance 
violation. Such constraints have important implications for 
quantum gravity models as well as models involving large extra dimensions. 
We also discuss the implications of observations of high energy \grays\ from 
the Crab Nebula on constraining quantum gravity models.

\end{abstract}


\end{frontmatter}

\section{Introduction}

In this paper, I will first present and discuss the evidence for absorption 
features at the multi-TeV end of the \gray\ spectra of the two active 
galactic nuclei for which we have the most detailed and highest quality
spectral data. These \gray\ sources are the BL Lac objects known as Markarian 
(Mkn) 501 and Mkn 421. I will then show how this absorption is due to the 
expected \gray\ annihilation caused by electron-positron pair production 
interactions of these \grays\ with intergalactic low energy photons having the 
flux level as determined by theoretical considerations combined with various 
astronomical observations. 

Stecker and Glashow have shown that the very existence of these 
interactions puts quantitative constraints on Lorentz invariance violation at 
high energies \cite{sg01}. I will show here that this also implies significant
constraints on a class of propsed models where quantum effects at the Planck
length scale of $\sim 10^{-35}$m, or a much larger modified Planck length 
scale in the case of large extra dimension models, can alter the relativistic 
energy-momentum dispersion relations for particles at energy scales much 
lower than the Planck scale ($\sim 10^{19}$ GeV). I will also discuss more
severe indirect constraints on quantum gravity models which follow from
observations of the \gray\ emission spectrum of the Crab Nebula.

\section{Absorption of Gamma-Rays at Low Redshifts}
%\section{The Opacity of Intergalactic Space Owing to the Cosmic Infrared
%Background Radiation}

The formulae relevant to absorption calculations involving pair-production 
are given and discussed in Ref. \cite{st92}.
For $\gamma$-rays in the TeV energy range, the pair-production cross section 
is maximized when the soft photon energy is in the infrared range. In terms
of the observed soft photon wavelength and \gray\ energy as a function of 
redshift, $z$,
\begin{equation} 
{\lambda\over{(1+z)}}~ \simeq~ \lambda_{e}{E_{\gamma}(1+z)\over{2m_{e}c^{2}}}~ =~ 1.24E_{\gamma,TeV}(1+z) \; \; \mu m 
\end{equation}
where $\lambda_{e} = h/(m_{e}c)$ 
is the Compton wavelength of the electron.
For a 1 TeV $\gamma$-ray emitted by a source at low redshifts, this 
corresponds to a soft photon having a
wavelength in the J-band of IR astronomy (1.25\mic). (Pair-production 
interactions actually take place with photons over a range of 
wavelengths around the optimal value as determined by the energy 
dependence of the cross section.) If the emission spectrum of
an extragalactic source extends beyond 20 TeV, then the extragalactic
infrared field should produce an absorption effect in the {\it observed} 
spectrum between $\sim20$ GeV and $\sim 5$ TeV, depending on the redshift of 
the source \cite{ss98}.

\begin{figure}
\centerline{\psfig{file=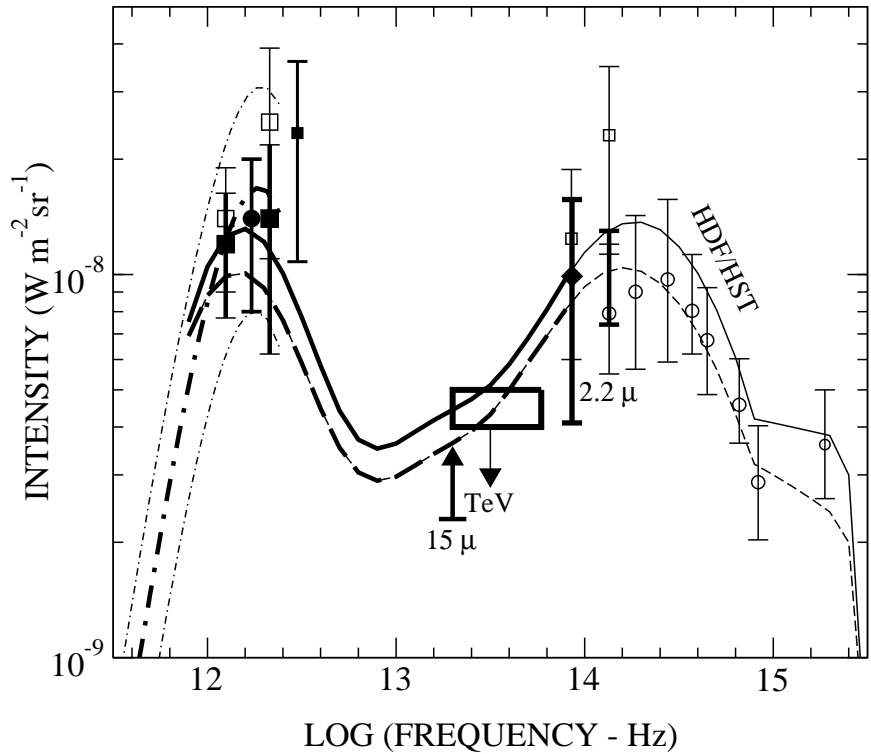,width=13.0truecm}}
\vspace{-8.0truecm}
\caption{The spectral energy distribution (SED) of the extragalactic infrared
background\protect\cite{ds02}. The data are given along with the two semi-
empirical model SEDs. The point at 100\mic\ indicated by the small square 
derived in Ref. \protect\cite{la00} is less certain than the data at lower 
frequencies because isotropy was not established for this point.} 
\label{IRSED}
%vspace{2.0truecm}
\end{figure}

Absorption of high energy \grays\ from extragalactic sources occurs {\it via} 
interactions of these photons with low energy photons of intergalactic 
radiation. These low energy photons are produced by stellar radiation and 
the reemission of such radiation by interstellar dust in galaxies. These 
photons then leave the galaxies in which they were produced, escaping into 
intergalactic space.

Malkan and Stecker (MS01) \cite{ms01} used empirically based spectral energy 
distributions (SEDs) of galaxies as a function of galaxy luminosity together 
with  galaxy luminosity distribution functions to derive the SED of 
the cosmic infrared background (CIB). The advantage of using empirical
data to construct the SED of the CIB, as done in MS01, is 
particularly important in the mid IR range. In this region of the spectrum,
galaxy observations indicate more flux from warm dust in galaxies than that
taken account of in some more theoretically oriented models, {\it e.g.,}
Ref. \cite{mp96}. De Jager and Stecker (DS02) \cite{ds02}
extended these SEDs into the optical and UV using a hybrid model based on 
Hubble Space Telescope (HST) galaxy counts from the Hubble Deep Field (HDF).
The lower curve in Figure \ref{IRSED} from DS02 (adapted from MS01) assumes a
redshift (temporal) luminosity evolution of galaxies  
$\propto (1+z)^3$ out to $z_{flat}=2$ (baseline model), whereas 
the upper curve assumes ``fast'' $\propto (1+z)^4$ evolution out to 
$z_{flat}=1.3$. For $z > z_{flat}$ no evolution is assumed to occur.
This is because evolution in star formation and stellar emissivity in galaxies
appears to level off at redshifts greater than $2$. 
\cite{mad98}-\cite{mo00} 
%so that the two curves in Figure \ref{IRSED} may be considered to be 
%reasonable lower and upper limits, bounding the expected IR flux.

Figure \ref{IRSED} shows the SED curves from SD02 in comparison with various 
data and limits.
The results of the MS01 \cite{ms01} fast evolution SED generally agree well 
with directly measured {\it COBE} (Cosmic Background Explorer) data.
These results are also in agreement with upper limits obtained from
TeV \gray\ studies \cite{dw94} - \cite{bil98} and lower limits obtained
from various infrared galaxy count studies.

\section{Observations of the Blazars Mkn 501 and Mkn 421}

The highest energy extragalactic \gray\ sources in the known universe are
the active galaxies called `blazars', objects that emit jets of relativistic 
plasma aimed directly at us.
Those blazars  known as X-ray selected BL Lac objects (XBLs), or
alternatively as high frequency BL Lac objects (HBLs), are expected to emit
photons in the multi-TeV energy range, but  only the nearest ones
are expected to be observable at TeV energies, the others being hidden by 
intergalactic absorption \cite{st96}.

Extragalactic photons with the highest energies yet observed originated 
in a powerful flare coming from the giant elliptical active galaxy known as 
Mkn 501 \cite{ah99}. Its spectrum is most easily understood and interpreted
as manifesting the high energy absorption to be expected from \gray\ 
annihilation by extragalactic pair production interactions.
The analyses of de Jager and Stecker (DS02) \cite{ds02}
and Konopelko {\it et al.} \cite{ko03} indicate the presence of the
absorption effect predicted by calculating the expected energy
dependent opacity inferred from the background light SEDs of of MS01 and DS02
(see previous section).This absorption is the result of electron-positron 
pair production by interactions of the multi-TeV \grays\ from Mkn 501 
primarily with intergalactic infrared photons.
Intrinsic absorption by pair production interactions within \gray\ sources 
such as Mkn 421 and Mkn 501 is expected to be negligible because such 
giant elliptical galaxies contain little dust to emit infrared radiation 
and because BL Lac objects have little gas (and therefore most likely 
little dust) in their nuclear regions. It also appears that \gray\ 
emission in blazars takes place at superluminal knots in 
their jets, downstream of the radio cores of these active galaxies and 
therefore downstream of any putative accretion disks \cite{jo01}.

\begin{figure}
\centerline{\psfig{file=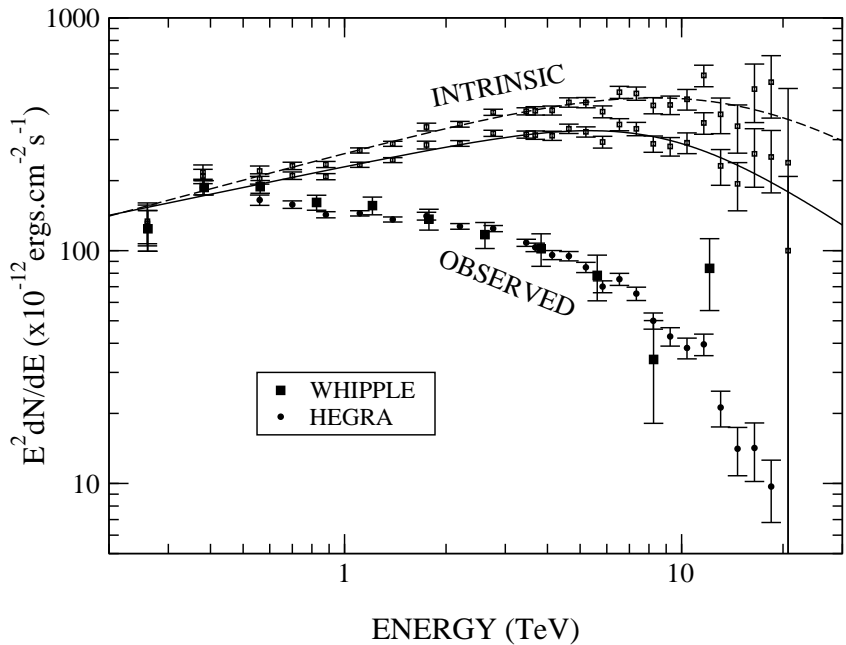,width=13.0truecm}}
\vspace{-8.0truecm}
\caption{The observed and derived intrinsic spectrum for Mkn 501 during the
April 1997 flare. 
The intrinsic spectra are given for the baseline and fast
evolution models shown in Figure \ref{IRSED} \protect\cite{ds02}.} 
\vspace{0.5truecm}
\label{dsfig}
\end{figure}

The spectrum of Mkn 501 in the flaring phase extends 
to an energy of at least 24 TeV \cite{ah99}. The DS02 calculations predict that
intergalactic absorption should strongly supress the spectra of these sources 
at multi-TeV energies. Figure \ref{dsfig} shows observed spectrum from 
{\it HEGRA} \cite{ah99} and the {\it Whipple} telescope \cite{kr99} 
(lower curve and points) and the derived intrinsic spectrum (upper 
curve and points) for Mkn501 in the flaring phase as given by de Jager 
and Stecker \cite{ds02}. The intrinsic spectrum was derived by correcting
for the opacity calculated for $z = 0.03$ as a function of energy, based on 
models of MS01, extended into the optical and UV range \cite{ds02}. 

\begin{figure}
\centerline{\psfig{file=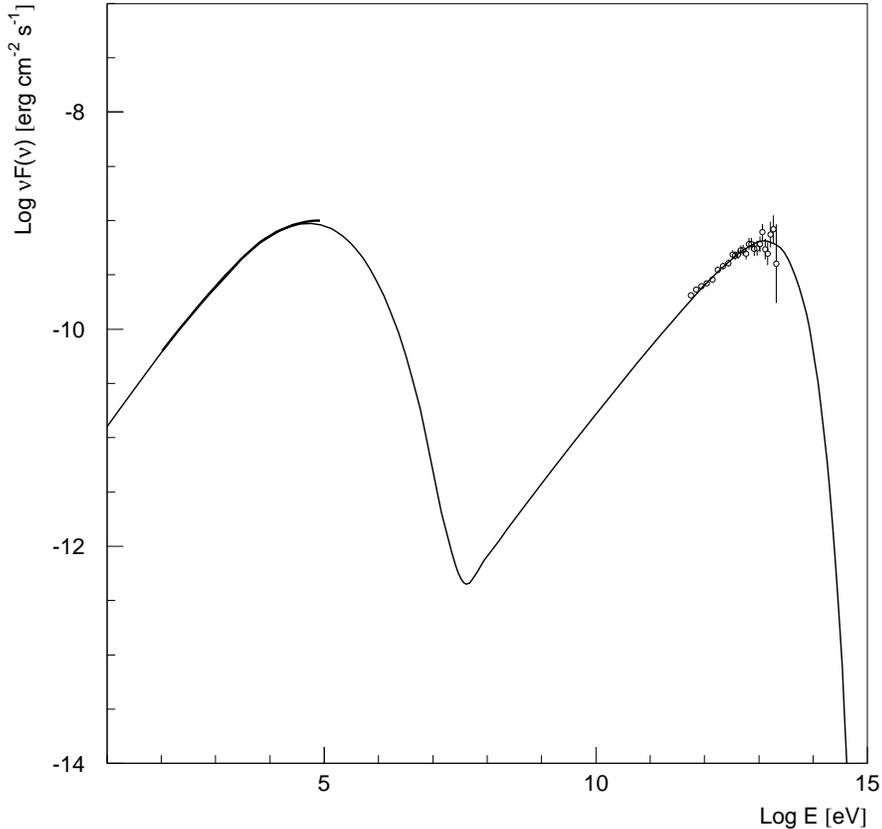,width=13.0truecm}}
%\vspace{-8.0truecm}
\caption{The April 1997 flaring spectrum of Mkn501 with absoprtion
taken into account at TeV energies, shown along with the SSC model fit
of Konopelko {\it et al.} \protect\cite{ko03} over a large energy range.} 
\vspace{1.0truecm}
\label{ko501}
\end{figure}

\begin{figure}
\centerline{\psfig{file=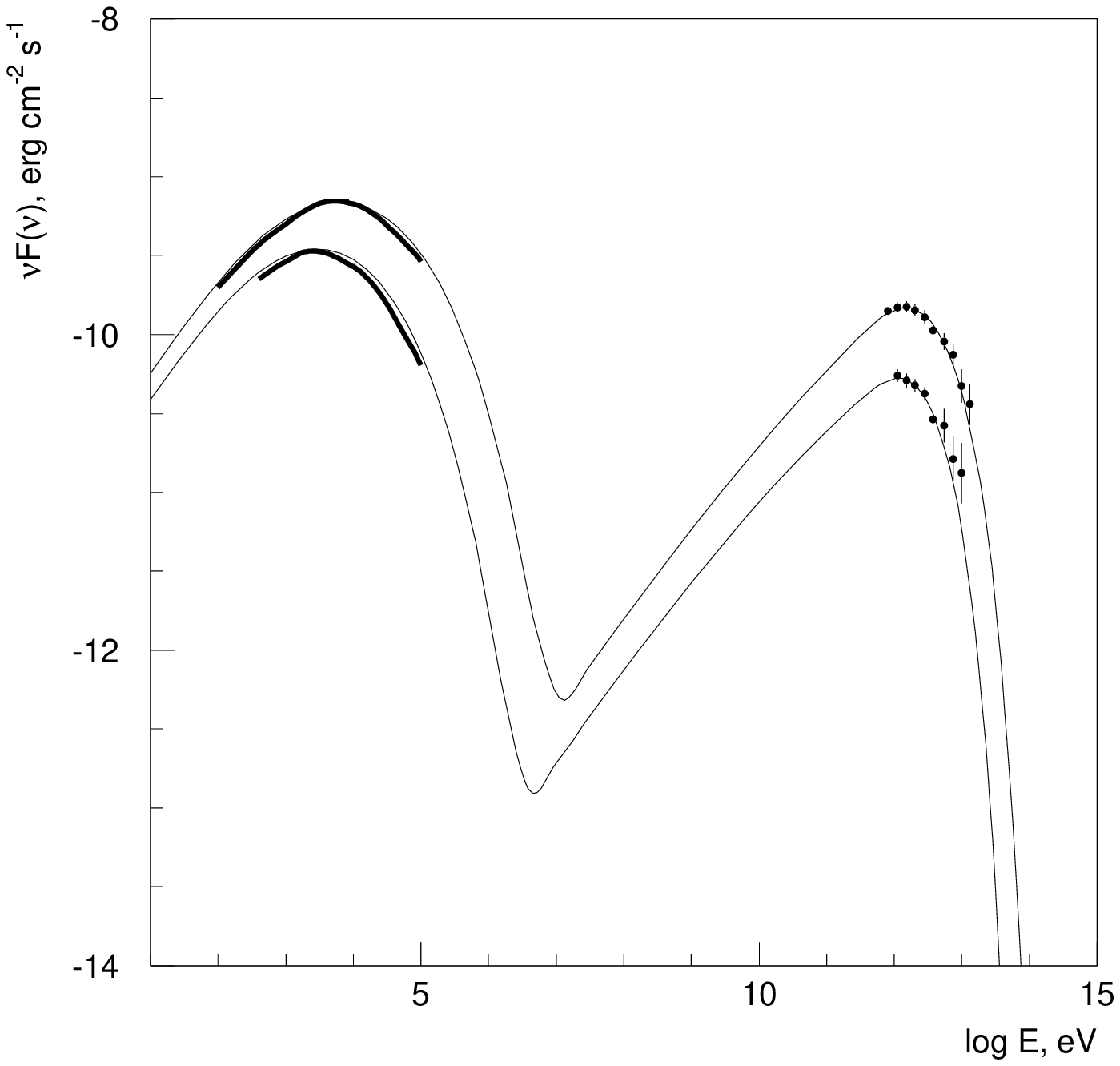,width=13.0truecm}}
%\vspace{-8.0truecm}
\caption{The spectra of Mkn421 in two phases with absoprtion
taken into account at TeV energies, shown along with the SSC model fit
of Konopelko {\it et al.} \protect\cite{ko03} over a large energy range.} 
\vspace{0.1truecm}
\label{ko421}
\end{figure}

Konopelko, {\it et al.} \cite{ko03} have reexmained the spectra of both Mkn 
421 and Mkn 501 corrected for absorption and have found that they fit  
synchrotron self-Compton (SSC) emission models from the radio to TeV 
\gray\ range. Both the X-ray and intrinsic TeV spectra of Mkn 421 peak at
lower energies than those of the flaring spectrum of Mkn 501 
(see Figs. \ref{ko501} and
\ref{ko421}) which can be understood if electrons were accelerated to higher
energies in the April 1997 flare of Mkn 501 than in Mkn 421.
As shown in Figure \ref{dsfig}, the intrinsic spectrum of Mkn 501 with
the absorption effect removed actually peaks at multi-TeV energies, rather than
falling off in this energy range. Thus, it appears that the dropoff in 
the observed \gray\ spectrum of Mkn 501 above $\sim$ 5 TeV is a direct 
consequence of intergalactic absorption. We will therefore interpret the Mkn 
501 data as evidence for intergalactic absorption with no indication of 
Lorentz invariance violation (see next section) up to a photon energy of 
$\sim\,$20~TeV. This conclusion can be substantiated by observations of other
extragalactic \gray\ sources at differing redshifts \cite{ss98}.

\section{High Energy Consequences of Breaking of Lorentz Invariance} 

It has been suggested that Lorentz invariance (LI) may be only an approximate 
symmetry of nature \cite{sa72} - \cite{co98}. A simple 
and self-consistent framework for analyzing possible departures from exact LI 
was suggested by Coleman and Glashow \cite{co99}, who assume LI to be broken
perturbatively in the context of conventional quantum field theory. Small
Lorentz noninvariant terms are introduced that are renormalizable, being
of mass dimension no greater than four and having dimensionless coupling
constants in the kinetic part of the Lagrangian. These terms are also chosen
to be gauge invariant under $SU(3)\times SU(2)\times U(1)$ 
(``the almost standard model''). It is further 
assumed that the Lagrangian is rotationally invariant in a preferred frame
which is presumed to be  the rest frame of the cosmic microwave background.

Consequent observable manifestations of LI breaking 
can be described quite simply in terms of different
maximal attainable velocities of different particle species as
measured in the preferred frame. This is because the small LI violating
terms modify the free-field propagators so that the maximum velocities  
of various particles are not equal to $c$. Indeed, this type of LI breaking 
within the hadron sector is one way to 
circumvent the predicted but unseen `GZK cutoff' in the ultrahigh energy 
cosmic-ray spectrum owing to photomeson interactions of such cosmic rays
with photons of the 2.7K cosmic
background radiation \cite{gr66}, \cite{zk66}.
These interactions are expected to produce an effective attenuation 
mean-free-path for ultrahigh energy cosmic rays ($E > 10^{20}$ eV) in 
intergalactic space of $< 100$ Mpc \cite{st68} in the absence of LI breaking.

\section{The LI Breaking Parameter $\delta$}

The discussion in the previous section shows that if LI is violated the 
maximum attainable velocity of an electron need not equal the velocity of 
light {\it in vacuo\/}, {\it i.e.,}  $c_e \ne c_\gamma$. The physical
consequences of this violation of LI depend on the sign of the difference 
between $c_e$ and $c_{\gamma}$ \cite{co99}. Following the discussion of 
Stecker and Glashow \cite{sg01}, we define

\begin{equation}
c_{e} \equiv c_{\gamma}(1 +  \delta) ~ , ~ ~~~0< |\delta| \ll 1\;,  
\end{equation}

\noindent
and consider the two cases of positive and negative values of $\delta$
separately. 

{\it Case I:} If $c_e<c_\gamma$ ($\delta < 0$), 
the decay of a photon into an electron-positron pair is kinematically allowed
for photons with energies exceeding

\begin{equation}
E_{\rm max}= m_e\,\sqrt{2/|\delta|}\;. 
\end{equation}

\noindent The  decay would take place rapidly, so that photons with energies 
exceeding $E_{\rm max}$ could not be observed either in the laboratory or as 
cosmic rays. Since photons have been observed with energies   
$E_{\gamma} \ge$ 50~TeV from the Crab nebula \cite{ta98}, we deduce for this 
case that $E_{\rm max}\ge 50\;$TeV, or that $|\delta| < 2\times  
10^{-16}$ \cite{sg01}. Stronger bounds on $\delta$ can be set
through observations of very high energy (TeV) photons.
The detection of cosmic $\gamma$-rays with energies
greater that 50~TeV from sources within our galaxy would improve the bound on
$\delta$.

{\it Case II:}  Here we are concerned with the remaining possibility, where
$c_e>c_\gamma$ ($\delta > 0$) and electrons become superluminal if their
energies exceed $E_{\rm max}/\sqrt{2}$.
Electrons traveling faster than light will emit light  at all frequencies by a
process of `vacuum \v{C}erenkov radiation.' This process occurs rapidly, so
that superluminal electron energies quickly approach $E_{\rm max}/\sqrt{2}$. 
Because electrons have been seen in the cosmic radiation 
with energies up to $\sim\,$2~TeV\cite{ni00}, it follows that 
$\delta <  3 \times 10^{-14}$. This upper limit is about two orders of
magnitude weaker than the limit obtained for Case I. 
  
\begin{figure}
\centerline{\psfig{file=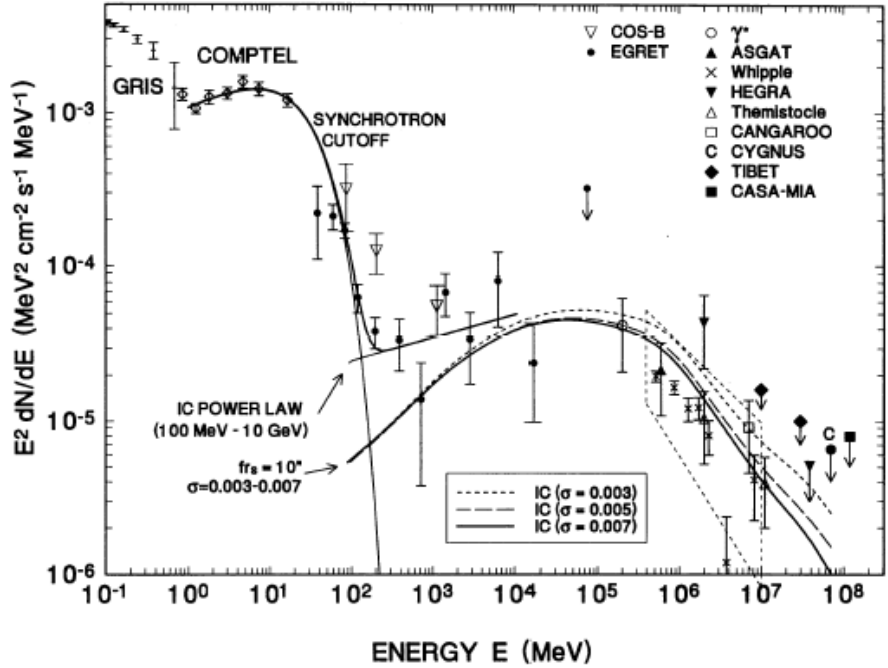,width=13.0truecm}}
%\vspace{-2.0truecm}
\caption{The observed \gray\ spectrum of the Crab nebula with curves showing 
the synchrotron and Compton components (from Ref. \protect\cite{de96}).}
\label{crab}
%\vspace{-1.0truecm}
\end{figure}

A smaller, but more indirect, upper limit on $\delta$ for the $\delta > 0$
case can be obtained from theoretical considerations of \gray\ emission from 
the Crab Nebula. The unpulsed \gray\ spectrum of the Crab Nebula can be 
understood to be produced by sychrotron emission up to the $\sim 0.1$ GeV 
\gray\ energy range (see Figure \ref{crab}).
Above 25 MeV, the synchrotron component falls off very rapidly with energy
as expected from theoretical limits on electron acceleration \cite{de96}. 
Emission above 0.1 GeV, extending into the TeV range, can be explained as
synchrotron self-Compton emission of the same relativistic electrons which
produce the synchrotron radiation. The Compton component, extending to 50
TeV, implies the existence of electrons having energies at least this great
in order to produce 50 TeV photons, even in the extreme Klein-Nishina limit.
This is, of course, required by conservation of energy.
This indirect argument, based on the reasonable assumption that the 50 TeV
\grays\ are from Compton interactions, leads to a smaller upper limit on 
$\delta$, {\it viz.,} $\delta < 10^{-16}$. Better observational \gray\ data 
for the Crab Nebula at multi-TeV energies, and data on the shape of its \gray\ 
spectrum above 50 TeV, will be needed to further test this indirect 
constraint on $\delta$. One open question involves the possibility of
hadronic interactions involving cosmic ray nucleons producing $\pi^0$'s 
which decay into observed high energy \grays (although there is no present 
indication of this). Hopefully, better data will determine whether or not
there is a significant hadronically induced component contributing to the 
multi-TeV spectrum of the Crab.

A further constraint on $\delta$ for $\delta > 0$ ($c_e>c_\gamma$)
follows from the modification of the threshold energy for the pair 
production process $\gamma + \gamma \rightarrow e^+ + e^-$. This follows from
the fact that the square of the four-momentum is changed to give the
threshold condition

\begin{equation}
2\epsilon E_{\gamma}(1 - \cos \theta) - 2E_{\gamma}^2\delta  >4 m_{e}^2,
\end{equation}

\noindent where $\epsilon$ is the energy of the low energy photon and $\theta$
is the angle between the two photons. The second term on the left-hand-side
comes from the fact that $c_{\gamma} =  
\partial E_{\gamma}/\partial p_{\gamma}$.

For head-on collisions ($\cos \theta = -1$) the minimum low energy photon
energy for pair production becomes 

\begin{equation}
\epsilon_{min} = {m_{e}^2\over{E_{\gamma}}} +  {E_{\gamma}\,\delta\over{2}}.
\end{equation}

It follows that the condition for a significant increase in the energy
threshold for pair production is $E_{\gamma}\delta/2$ $ \ge$
$ m_{e}^2/E_{\gamma}$, or 
equivalently, 

\begin{equation}
\delta \ge {2m_{e}^{2}\over{E_{\gamma}^{2}}}.
\end{equation}

As discussed in the previous section, there is no indication of LI violation
supressing the physics of pair production for photons up to an energy of
$\sim20$ TeV. Thus, it follows from eq. (6) that the Mkn 501 observations
imply the constraint $\delta \le 2m_{e}^{2}/E_{\gamma}^{2} = 
1.3 \times 10^{-15}$ \cite{sg01}. This constraint on positive $\delta$ is 
more secure than the smaller, but indirect, limit given by the Crab Nebula 
acceleration model.

\section{Quantum Gravity Models}

In the absence of a true and complete theory of quantum gravity, theorists 
have been suggesting and exploring models to provide experimental and 
observational tests of possible manifestations of quantum gravity phenomena. 
Such phenomena have usually been suggested to be a possible result of quantum 
fluctuations on the Planck scale $\mpl = \sqrt{\hbar c/G} \simeq 1.22 
\times 10^{19}$ GeV/c$^2$, corresponding to a length scale $\sim 1.6 \times 
10^{-35}$ m \cite{ga95} - \cite{al02}. In models involving large extra 
dimensions, the energy scale at which gravity becomes strong can be much 
smaller than $\mpl$, with the quantum gravity scale, $\mqg$, approaching the 
TeV scale \cite{el00}, \cite{el01}.

In many of these models Lorentz invariance is predicted to be violated at high 
energy. This results in interesting modifications of particle physics that 
are accesible to observational tests using TeV \gray\ telescopes and cosmic 
ray detectors. An example of such a model is the loop quantum gravity model 
with a preferred inertial frame given by the cosmological rest frame of the 
cosmic microwave background radiation (For an extensive discussion, see the
review given in Ref. \cite{sm03}.) 

In the most commonly considered of these models, the usual relativistic 
dispersion relations between energy and momentum of the photon and the electron

\begin{equation}
E_{\gamma}^2 = p_{\gamma}^2 
\end{equation}

\begin{equation}
E_{e}^2 = p_{e}^2 + m_{e}^2
\end{equation}

(with the ``low energy'' speed of light, $c \equiv 1$) are modified by a 
leading order quantum space-time geometry corrections which
are cubic in $p \simeq E$ and are supressed by the quantum gravity mass 
scale $M_{QG}$. Following Refs. \cite{am98} and \cite{al02}, we take the 
modified dispersion relations to be of the form

\begin{equation}
E_{\gamma}^2~ =~ p_{\gamma}^2~ -~ {p_{\gamma}{^3}\over M_{QG}} 
\end{equation}

\begin{equation}
E_{e}^2~ =~ p_{e}^2~ +~ m_{e}^2 ~-~ {p_{e}{^3}\over M_{QG}} 
\end{equation}

We assume that the cubic 
terms are the same for the photon and electron as in eqs. (9) and (10). More 
general formulations have been considered by Jacobson, Liberati and 
Mattingly \cite{jlm02} and Konopka and Major \cite{ko02}. 

As opposed to the Coleman-Glashow formalism, which involves mass dimension four
operators in the Lagrangian and preserves power-counting renormalizablility,
the cubic term which modifies the dispersion relations may be considered in 
the context of an effective 
``low energy'' field theory, valid for $E \ll M_{QG}$, in which case the 
cubic term is a small perturbation involving dimension five operators 
whose construction is discussed in Ref. \cite{my03}. With this caveat, we can 
generalize the LI violation parameter $\delta$ to an energy dependent form

\begin{equation}
\delta~ \equiv~ {\partial E_{e}\over{\partial p_{e}}}~ -~ {\partial E_{\gamma}
\over{\partial p_{\gamma}}}~
 \simeq~ {E_{\gamma}\over{M_{QG}}}~ 
-~{m_{e}^{2}\over{2E_{e}^{2}}}~ -~ {E_{e}\over{M_{QG}}} ,
\end{equation}

which is a valid approximation for the energy regime $E_{e} \gg m_{e}$.
Note that the maximum velocities of particles of type $i$ are reduced
by ${\cal{O}}(E_{i}/M_{QG})$.

For pair production then, with the positron and electron energy 
$E_{e} \simeq E_{\gamma}/2$,

\begin{equation}
\delta~ =~ {E_{\gamma}\over{2M_{QG}}}~ -~ {2m_{e}^{2}\over{E_{\gamma}^{2}}} 
\end{equation}

and the threshold condition given by eq.(6) reduces to the constraint

\begin{equation}
M_{QG} ~\ge~ {E_{\gamma}^3\over 8m_{e}^2}.
\end{equation}

Since pair production occurs for energies of at least 20 TeV, as indicated
by our analyses of the Mkn 501 and Mkn 421 spectra \cite{ds02},\cite{ko03}, 
we then find the constraint on the quantum gravity scale
$M_{QG} \ge 0.3 M_{Planck}$. This constriant contradicts the 
predictions of some proposed quantum gravity models involving large extra 
dimensions and smaller effective Planck masses. Previous constraints on 
$M_{\rm QG}$ for the cubic model, obtained from limits on the energy 
dependent velocity dispersion of \grays\ for a TeV flare in Mkn 421 
\cite{bi99} and from \gray\ bursts \cite{sc99} were in the much less 
restrictive range $M_{QG} \ge (5-7) \times 10^{-3} M_{Planck}$.

\section{Beyond the Planck Scale: Implied Constraints from Crab Nebula \grays}

Within the context of a more general cubic modification of the dispersion 
relations given by eqs. (9) and (10), Jacobson, {\it et al.} \cite{jlm03} 
have obtained an indirect limit on $M_{QG}$ from the apparent cutoff in the 
synchrotron component of the in the Crab Nebula \gray\ emission at $\sim 0.1$ 
GeV (see Figure \ref{crab}). 
By making reasonable assumptions to modify the standard synchrotron
radiation formula to allow for Lorentz invariance violation, they have 
concluded that the maximium synchrotron photon energy will be given by 
$E_{\gamma, {\rm max}}$ = 0.34 $(eB/m_{e})(m_{e}/M_{QG})^{-2/3}.$
This reasoning leads to the constraint $ M_{QG}$$ >$$ 1.2 \times 10^{7} 
M_{Planck}.$

Future observations of the Crab Nebula with the {\it GLAST} (Gamma-Ray 
Large Area Space Telescope) satellite, scheduled to be launched in 2005,
will provide a better determination of its unpulsed \gray\ spectrum 
in the energy range above 30 MeV where the transition from the synchrotron
emission component to the Compton emission component occurs. This will
provide a more precise determination of the maximum electron energy in the
Nebula and therefore provide a more precise constraint on the parameter
$M_{QG}$ as we have defined it here. However,
this constraint will still be orders of magnitude above the Planck scale.

\section{Conclusions}

Nearly a century after the inception of special relativity, high energy 
\gray\ observations have confirmed its validity up to electron energies
of 2 TeV, photon energies of 20 TeV and, indirectly, up to electron energies
in the PeV range. These results indicate an absence of evidence for 
proposed violations of Lorentz invariance as predicted by some 
phenomenological quantum 
gravity and large extra dimension models. Thus, high energy astrophysics 
has provided important empirical constraints on Planck scale physics.

Models with large extra dimensions are ruled out by the existence of 
absorption in the very high energy spectra of nearby BL Lac objects. 
The fact that more distant brighter sources are not seen can also be
taken as indirect evidence of intergalactic absorption by pair production
interactions \cite{st96}.

The constraints based on analysis of the Crab Nebula \gray\ spectrum,
discussed in the previous section, imply that the quantum
gravity scale is orders of magnitude above the Planck mass scale. This
indicates that the class of models considered here cannot be
reflective of physics at the Planck scale. Models such as loop quantum gravity 
with a preferred inertial frame are ruled out by this line of reasoning. 
Alternative models to consider might be models with a 
quartic term with $M_{QG}^2$ supression in the dispersion relations, 
Lorentz invariant quantum gravity models, or really new Planck scale 
physics such as string theory, which preserves Lorentz invariance.

\section*{Acknowledgments}
I would like to thank Ted Jacobson, Stefano Liberati, David Mattingly and
Serge Rudaz for helpful discussions.

\end{document}